\documentclass{aa}  
\usepackage{graphicx}
\usepackage{txfonts}
\usepackage{natbib}
\begin{document} 

\title{Two epoch spectra-imagery of PV~Cep outflow system} 

   \author{T.A. Movsessian
          \inst{1}
          \and
          T.Yu. Magakian
          \inst{1}
          \and          A.V. Moiseev
          \inst{2}
                    \inst{}
          }
   \institute{Byurakan Astrophysical Observatory NAS Armenia, Byurakan, Aragatsotn prov., 0213,
              Armenia\\
                \email{tigmov@bao.sci.am; tigmag@sci.am}
          \and
              Special Astrophysical Observatory,
             N.Arkhyz, Karachaevo-Cherkesia, 369167 Russia\\
             \email{moisav@gmail.com}
                        }
   \date{Received ...; / Accepted ... }
%\thanks{Based on observations collected with the 6m telescope of the Special
%   Astrophysical Observatory (SAO) of the Russian Academy of Sciences (RAS),
%   operated under the financial support of the Science Department of Russia
%   (registration number 01-43.)}

   \date{Received ...; accepted ...}

% \abstract{}{}{}{}{} 
% 5 {} token are mandatory
 
  \abstract
   {We continue to study the structure and kinematics of HH flows. Herbig-Haro (HH) flows exhibit large variety of morphological and kinematical structures. Both proper motion (PM) and radial velocity investigations are essential to understand the physical nature of such  structures.}
{ We investigate the kinematics and PM of spectrally separated structures in the PV Cep HH flow HH 215.  }
   {We present the observational results obtained with a 6 m telescope (Russia) using the SCORPIO multi-mode focal reducer with scanning Fabry-Perot interferometer. Two epochs of the observations of the PV Cep region in H$\alpha$ and [SII] emission (2003 and 2020-2021) allowed us to study the morphology of HH 215 jet in detail and to measure the PM and the radial velocities for its inner structures.}
   {Already known emission knots in the HH 215 flow and new features were studied. Moreover, a newly-formed HH knot was revealed, presumably formed during the large maximum of PV Cep star in 1976-1977. 

We found the high-velocity inner channel in the HH 215 ionized outflow, oriented accordingly to the mean direction of the whole HH outflow and the axis of the symmetry of the reflection nebula. The HH-knots located along the axis of the high-velocity channel have a position angle coinciding with its axis (abut 325$^{\circ}$), however other ones have completely different value (about 25$^{\circ}$), which supports the idea that those knots are formed by oblique shocks. We derived the value of   i $\approx$ 30\degr$\pm$ 5\degr\  for the inclination angle between the flow axis and the line of sight.   
The total length of HH 215 outflow should be about 0.2 pc, and the full length of the bipolar outflow from PV\ Cep (HH 315 + HH 215) can be estimated as  3.6 pc, assuming that it more or less keeps the same inclination angle.

 }
   {}
   {}


\titlerunning{PV Cep jet}
  \authorrunning{Movsessian et al.}

   \maketitle
%
%-------------------------------------------------------------------
%_______________________________________________________________________________

%%%%%%%%%%%%%%%%%%%%%%%%%%%%%%%%%%%%%%%%%%%%%%%%%%%%%%%%%%%%%%%%%%%%%%%%%%%%%%%%
\section{Introduction\\ }
%%%%%%%%%%%%%%%%%%%%%%%%%%%%%%%%%%%%%%%%%%%%%%%%%%%%%%%%%%%%%%%%%%%%%%%%%%%%%%%%
\object{PV~Cep} star is one of the most remarkable eruptive pre-main-sequence objects. It is located in the northeastern edge of the \object{L~1158} and \object{L~1155} groups of dark clouds, in 1.5\degr\ to the east from the famous \object{NGC~7023} nebula. Its distance can be obtained from GAIA DR3 catalogue parallax:  356 pc. This value can be compared with the quite close estimate (343 pc) from the  work of \citet{Vioque}, based on the parallax from GAIA DR2.  For simplicity we will assume 350 pc as a distance to PV~Cep in the further analysis. This distance associate this object with the Cepheus Flare star forming region \citep{Kunetal}. 

This object was discovered in the period of its most powerful outburst in 1976  by \citet{GMA1977} and \citet{Cohen1977}, when its brightness raised for 5 magnitudes in red \citep{Cohen1981}. Since then this star underwent several subsequent maxima of various amplitude and duration; several previous outbursts which went unnoticed, were found  (Kun et al., unpublished; \citealt{AMMM2021}). 

Most interesting is the association of PV Cep  with  highly variable in the brightness and morphology reflection nebula of roughly conical shape. This nebula is known as \object{GM~1-29} \citep{GM1977} or \object{RNO~125} \citep{Cohen1980}. Deep CCD images of this object reveal a faint red “``counterfan'', proving its bipolar structure \citep{Levreault1987}. Recent changes in the shape of the nebula are shown by \citet{Kunetal2011}.

The highly variable spectrum of PV~Cep corresponds to very active T Tau star and contains a great amount of permitted and forbidden emission lines; several of them are split to many components. It was studied in detail in many works \citep[see, eg.][]{caratti,giannini}. PV~Cep was assigned to the EXor class, introduced by \citet{Herbig1977,Herbig1989}, but later was eliminated by him from the list   since further studies revealed too
many significant differences between PV~Cep and typical EXors. Several authors relate this object to the eruptive stars, intermediate between FUors and EXors \citep{AMMM2021}.

 Like the majority of PMS stars with eruptive activity, PV Cep was found to be the source of directed outflows. Its bipolar molecular outflow was found by \citet{Levreault1984}. The blueshifted
lobe of this outflow coincides with the axis of the reflection nebula while the redshifted lobe is associated with the faint ``counterfan''. PV~Cep has been also detected as a source of cm radio continuum  \citep{Anglada1992}, which, without any doubt, comes from the accretion disk near the star.

Soon the association of PV~Cep with an ionized Herbig-Haro (HH) outflow also was found.
\citet{Neckel1987} detected via spectral observations two HH emission knots, projected on the
reflection nebula. One of them was also found on the narrow-band
images \citep{Ray1987}. These same knots were afterwards rediscovered and described by  \citet{Reipurth1997} and \citet{Gomez1997}. Now they are known as \object{HH~215}. Besides, the existence of the giant (i.e. parsec-sized)
bipolar HH outflow from PV~Cep, which was numbered as \object{HH~315}, was revealed \citep{Gomez1997, Reipurth1997}. It exhibits a clear S-shaped point symmetry of blue
and red lobes suggesting precession of the source \citep{Reipurth1997}. All knots of this outflow
lie within the lobes of the
molecular outflow. 

Though the PV~Cep system still is a subject of many studies, the structure and kinematics of its collimated outflow, especially near the star (i.e. HH~215), where the influence of the bright reflection nebula is large, remained little studied.    In this paper we focused on nearby outflow system.

Our observations were carried out with Fabry-P\'erot scanning interferometry (FPI) technique in the H$\alpha$ and [\ion{S}{ii}]\,6716\AA\ emission lines. The HH~215 region was fully covered. Similar to
other 3D methods, FP scanning interferometry provides detailed
spectrophotometric information alongside with full spatial coverage,
but the spectral range is usually limited to one spectral line.
However, its distinct advantage is a wide field of view, combined
with high spectral resolution, ideal for studying extended emission objects such as Herbig-Haro (HH) jets.
To obtain the full three-dimensional picture of the jet kinematics, we also measured proper motions of all knots, where such
measurements were possible.

%%%%%%%%%%%%%%%%%%%%%%%%%%%%%%%%%%%%%%%%%%%%%%%%%%%%%%%%%%%%%%%%%%%%%%%%%%%%%%%%
\section{Observations and Data Reduction}
%%%%%%%%%%%%%%%%%%%%%%%%%%%%%%%%%%%%%%%%%%%%%%%%%%%%%%%%%%%%%%%%%%%%%%%%%%%%%%%%

Observations were carried out in the prime focus of the           6 m telescope of Special Astrophysical Observatory of the Russian Academy of Sciences in two epoches: 25 May 2003 and      29 Dec  2021 (in [\ion{S}{ii}]\,6716\AA\ emission) and 08 Dec  2020 (in H$\alpha$ \ emission), in good conditions (seeing was about  2\arcsec). The scanning FPI was placed in the parallel beam of the SCORPIO (in 2003) and SCORPIO-2 (in 2020-21) multi-mode focal reducers. These devices are described in \citet{AM2005} and \citet{AM2011}, their capabilities in the scanning FPI observational mode are reviewed in  \citet{moisav2021}.

                                                                         The detector used during the first epoch of observations was a EEV 42-40 2048$\times$2048 pixel CCD array. Observations were performed with 4$\times$4 pixel binning to reduce the readout time and readout noises, so 512$\times$512 pixel images were obtained in each spectral channel. The field of view was 5.3\arcmin\ and the scale 0.62$\arcsec$ per pixel.

The scanning interferometer, used in these observations, was Queensgate ET-50, operating in the 501st order of interference at the H$\alpha$ wavelength, providing spectral resolution of FWHM$\approx$ 0.8\AA\ (or $\approx$40 km s$^{-1}$) for a range of $\Delta\lambda$=13\AA\ (or $\approx$590 km s$^{-1}$) free from orders overlapping. The number of spectral channels was 36 and the size of a single channel was $\Delta\lambda\approx$ 0.36\AA\ ($\approx$16 km s$^{-1}$).  Exposure time was 200 seconds per channel, and total exposure time was 7200 sec. 

During the second epoch observations in 2020 with SCORPIO-2 the detector was a E2V 42-90 $4612\times2048$ pixel CCD array. Observations were performed with 2$\times2$ pixel binning in   a central square of the CCD array, so 1024$\times$1024 pixel images were obtained in each spectral channel. The field of view was 6.1\arcmin\ for a scale of 0.36\arcsec\ per pixel. Total exposure was 7200 seconds. 

In 2021  the detector was a E2V 261-84 $4096\times$2048 pixel CCD array.    Observations were performed with 2$\times2$ pixel binning in   a central square of the CCD array, so 1024$\times1024$ pixel images were obtained in each spectral channel. The field of view was 6.4\arcmin\ for a scale of 0.4\arcsec\ per pixel. Total exposure was 6000 seconds.                                                                                     

\begin{figure}
\includegraphics[width=20pc]{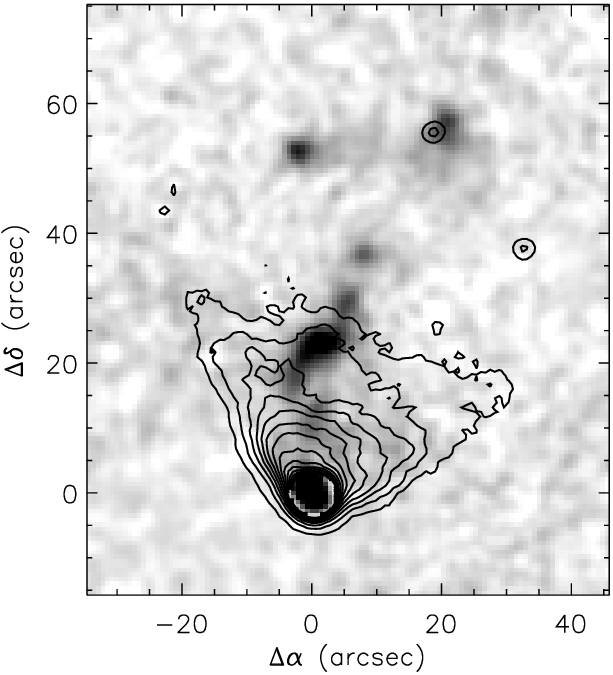}
\caption{ Restored image of the PV~Cep jet obtained from 2003 observations with scanning FPI: continuum subtracted [\ion{S}{ii}]\,6716\AA\ emission (gray scale) with superposed continual image (isolines). }
\label{fig1}
\end{figure}

The both second epoch SCORPIO-2 observations were performed with the ICOS scanning FPI  operating in the 751st order of interference at the H$\alpha$ wavelength, providing spectral resolution of FWHM $\approx$ 0.4\AA\ (or $\approx$20 km s$^{-1}$) for a range of $\Delta\lambda$=8.7\AA\ (or $\approx$390 km s$^{-1}$) free from order overlapping. The number of spectral channels was 40 and the size of a single channel was $\Delta\lambda\approx$ 0.22\AA\ ($\approx$10 km s$^{-1}$). \

In the first epoch only the observations in [\ion{S}{ii}]\,6716\AA\ line were carried out; in the second epoch both [\ion{S}{ii}]\,6716\AA\ and H$\alpha$ emissions were observed. In all cases the interference filters with FWHM = 15\AA,\ centered on the H$\alpha$ and [\ion{S}{ii}]\,6716\AA\ lines, were used for pre-monochromatization.
 
\begin{figure*}
\includegraphics[width=14.2pc]{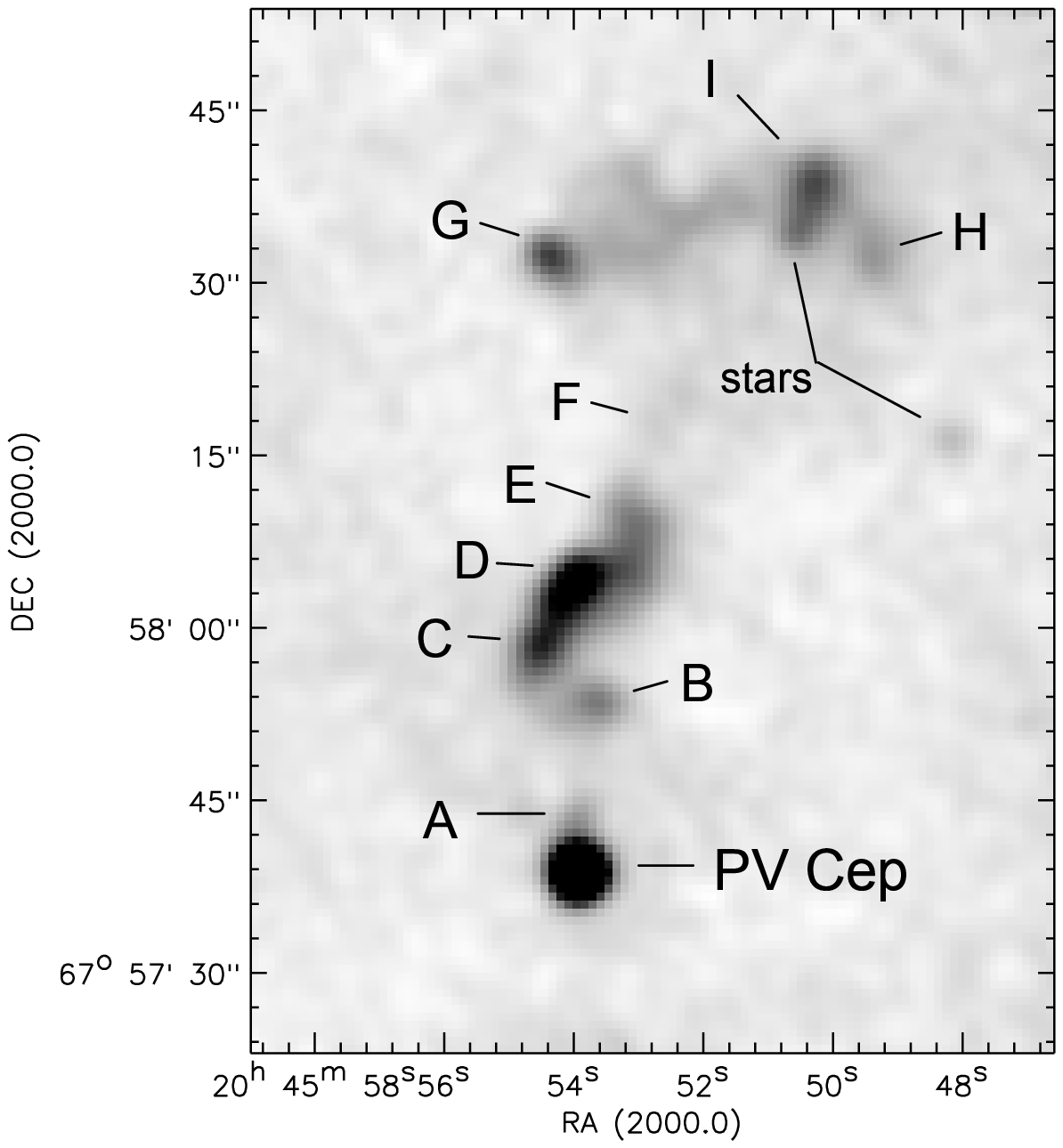} 
\includegraphics[width=15pc]{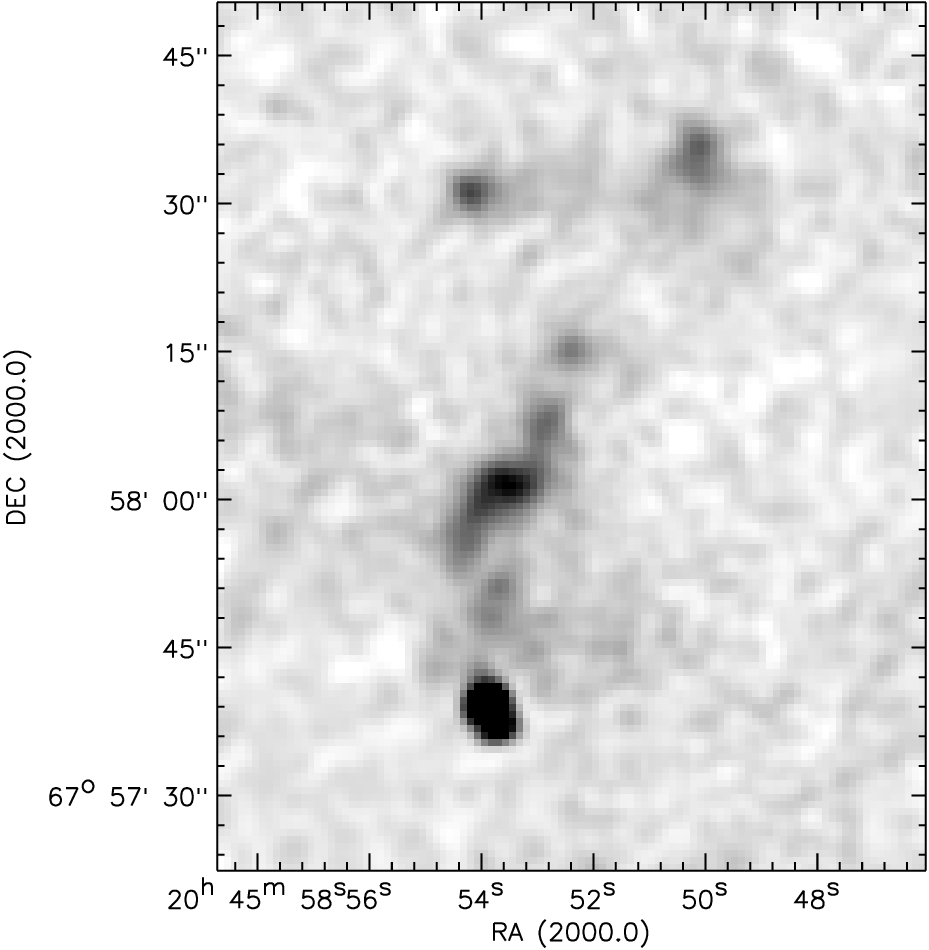} 
\includegraphics[width=13.7pc]{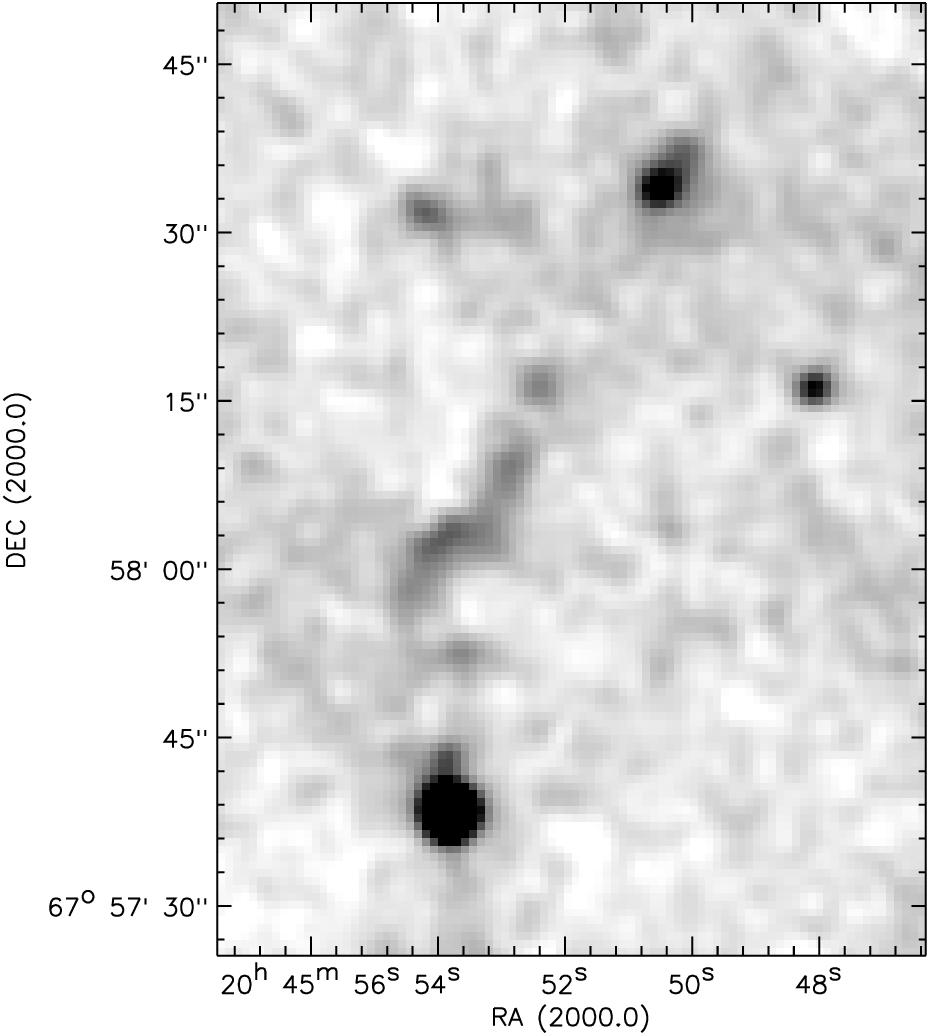} 
\caption{ Restored image obtained from 2020 observations with scanning FPI of the PV~Cep outflow in H$\alpha$ emission (left panel), as well as continuum subtracted [\ion{S}{ii}]\,6716\AA\  images restored from FPI\ observations in 2003 (central panel) and in 2020 (right panel). }
\label{fig2}
\end{figure*}

We reduced our interferometric observations using the software developed at the SAO \citep{moisav2008,moisav2021} and the ADHOC software package{\footnote{The ADHOC software package was developed by J. Boulestex (Marseilles Observatory) and is publicly available in the Internet.}}. After primary data reduction, subtraction of night-sky lines, photometric correction and wavelength calibration, the observational material represents ``data cubes''. Final data cubes were subjected to   optimal data filtering, which included Gaussian smoothing over the spectral coordinate with FWHM = 1.5 channels and spatial smoothing by a two-dimensional Gaussian with FWHM = 2--3 pixels.

Using these data cubes, we spectrally separated  the
structures in the outflow system. Then PM were measured for  the selected structures using observations in both epochs. For the PM estimation a
method of two images optimal offset computation by
means of cross-correlation was used (realized as  procedure CORREL\_IMAGES by F. Varosi and included in IDL astronomy library\footnote{\url{https://asd.gsfc.nasa.gov/archive/idlastro}; last version \url{https://github.com/wlandsman/IDLAstro}}).

%%%%%%%%%%%%%%%%%%%%%%%%%%%%%%%%%%%%%%%%%%%%%%%%%%%%%%%%%%%%%%%%%%%%%%%%%%%%%%%%
\section{Results}
%%%%%%%%%%%%%%%%%%%%%%%%%%%%%%%%%%%%%%%%%%%%%%%%%%%%%%%%%%%%%%%%%%%%%%%%%%%%%%%%

%%%%%%%%%%%%%%%%%%%%%%%%%%%%%%%%%%%%%%
\subsection{Morphology of the outflow}
%%%%%%%%%%%%%%%%%%%%%%%%%%%%%%%%%%%%%%

  Our observations cover a field of view of about 6\arcmin, including only the knots of HH~215 system, which is the initial  part of the giant outflow HH~315. The detailed description of the morphology of HH~215 still does not exist.The lack of such investigations is related to the difficulties of detecting of the faint emission structures against the bright background of the variable reflection nebula near PV~Cep. 

As was mentioned above, in 2003, during the observations with the Fabry-P\'erot interferometer, we scanned only the [\ion{S}{ii}]\,6716\AA\  line, because the reflection nebula was very bright in both the continuum and   the H$\alpha$ emission, due to the maximal brightness of the source star.  During the second epoch observations, PV~Cep was at minimum of brightness, and its reflection nebula was only barely visible. Taking into account these circumstances, we decided to perform the scanning of both the [\ion{S}{ii}]\,6716\AA\  and  H$\alpha$ lines.

Fig.\ref{fig1} presents the superposition of integrated images in the monochromatic [\ion{S}{ii}]\,6716\AA\  line emission (gray scale) and continuum (contours), restored from FPI observations in 2003. The cone-shaped morphology of the reflection nebula, as well as the wiggling structure of the emission jet, are clearly visible.  

 Second epoch observations in H$\alpha$ (Fig.\ref{fig2}, left panel) revealed the complex morphology
of HH~215 flow, which consists of the several well defined knots. As usual, we denote them by
letters, starting from the source star. Knot D can be identified with the first emission patch (P1),
mentioned by \citet{Neckel1987}, and knot G -- with the second one (P2), also detected by
\citet{Ray1987}. The correspondence between our knots and those  listed by \citet{Gomez1997}, is the following: HH 215(1) is the knot G, HH 215(2) - C, HH 215(3) and HH 215(4) - D, and HH 215(5) is probably the knot E.    In general, HH 215 flow can be described as the wiggling jet near the source (knots
A-F), which is advanced by the large arcuate structure (knots G-I). On the same figure, the images
obtained in two epochs in [\ion{S}{ii}]\,6716\AA\ emission are also presented (central and right
panels). Especially noteworthy is the appearance of a new knot A at a distance of about  4\arcsec \
from the source, which became visible only in the second epoch images. This knot is significantly
brighter in [\ion{S}{ii}]\,6716\AA\ than in H$\alpha$ emission, but is well discernible in both
wavelengths.
Besides, yet another conspicuously bright knot in [\ion{S}{ii}]\,6716\AA\  (marked as F),
which is only barely visible on the H$\alpha$ image,  should be noted.

%%%%%%%%%%%%%%%%%%%%%%%%%%%%%%%%%%%%%
\subsection{Radial velocities }
%%%%%%%%%%%%%%%%%%%%%%%%%%%%%%%%%%%%
  
\begin{figure}
\includegraphics[width=20pc]{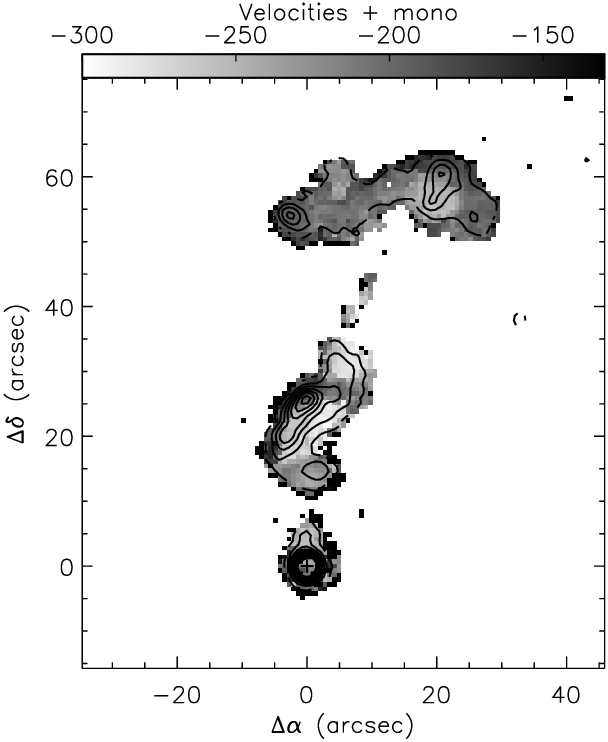}
\caption{ Two dimensional map of radial velocities in PV~Cep outflow system  obtained from 2003 FPI observations  (grey scale) with superposed monochromatic image in H$\alpha$ emission (isolines). }
\label{fig3}
\end{figure}

The differences in the measured radial velocities between the two epoch observations do not exceed  5 km s$^{-1}$ for knot D and are lower than 2 km s$^{-1}$ for other knots. Because of the higher spectral resolution, we will concentrate on the radial velocities, obtained in the second epoch; in any case,   the average radial velocities in the PV~Cep outflow are very large and vary around $-$300 km s$^{-1}$. Here and below by ``high'' and ``low'' velocities we  mean their absolute values.
Besides, here and through all this work
we use heliocentric radial velocities.

\begin{figure*}
\includegraphics[width=10.6pc]{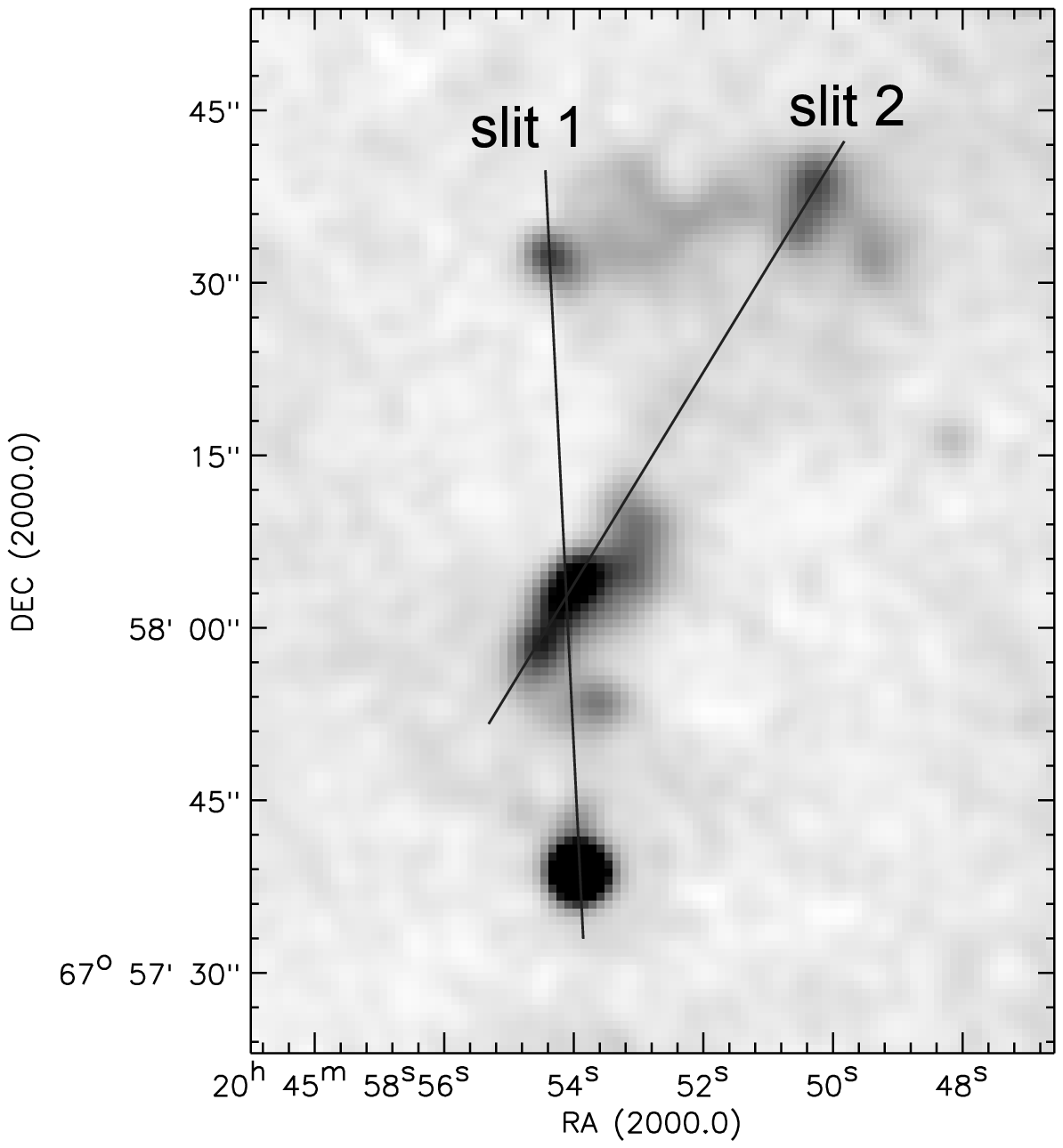} 
\includegraphics[width=12pc, angle=0]{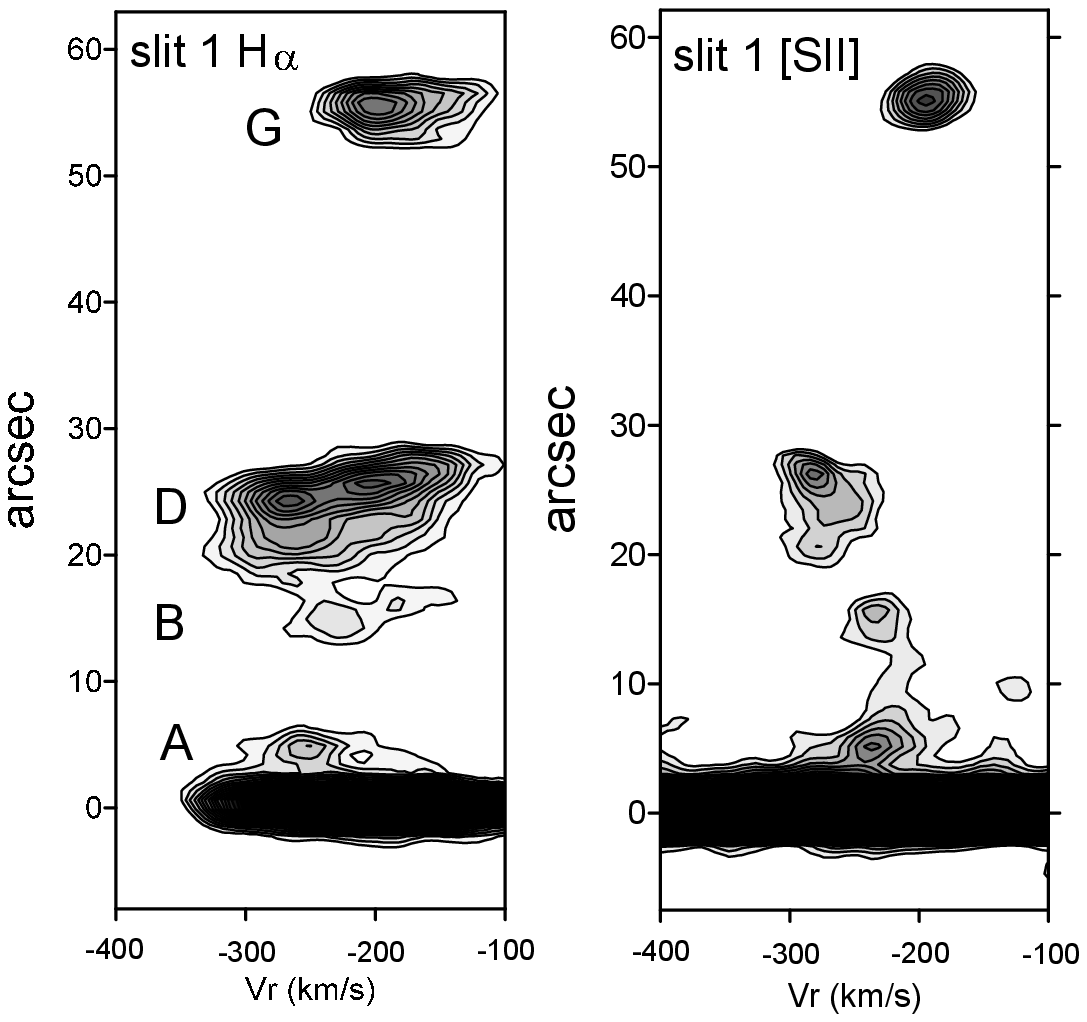} 
\includegraphics[width=7.2pc, angle=0]{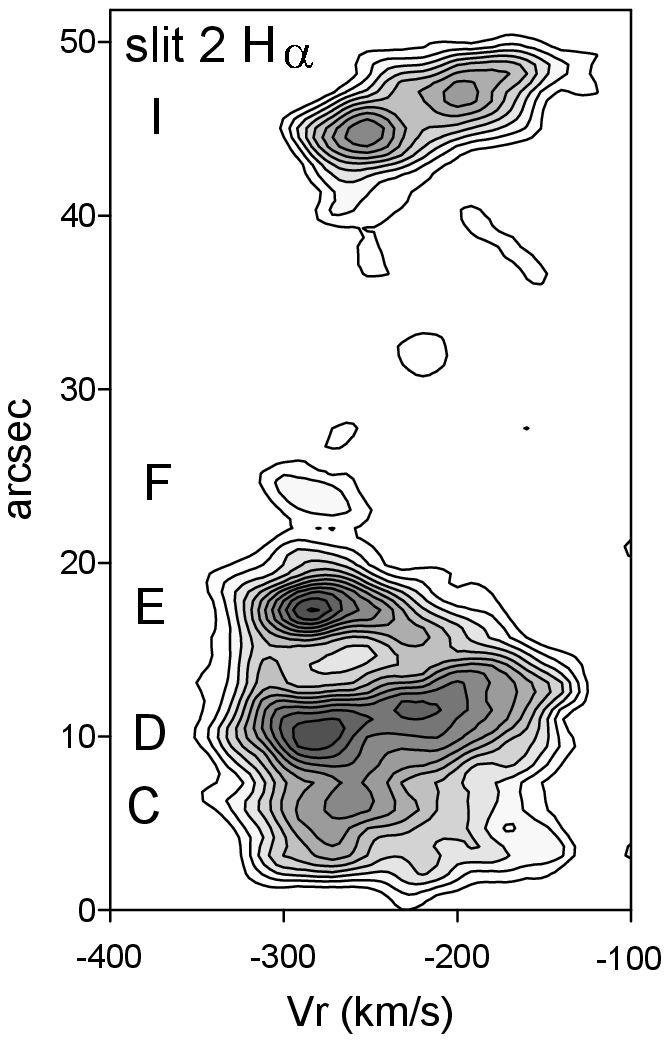}
\includegraphics[width=7.3pc, angle=0]{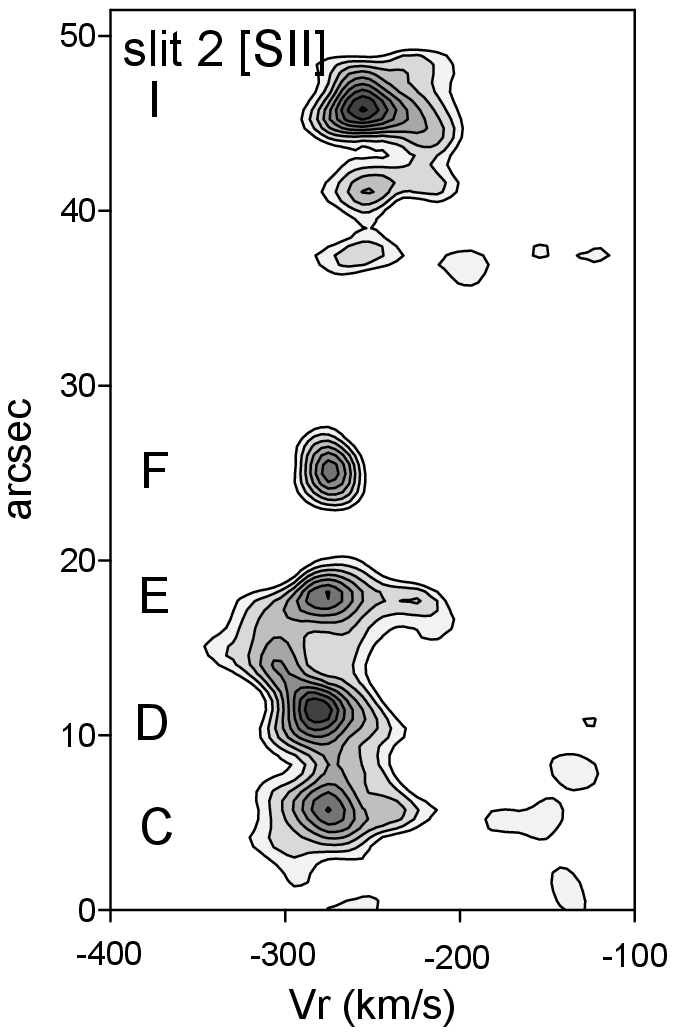}
\caption{ Positions of two pseudo-slits of marked on H$\alpha$ image of the PV\ Cep outflow system  (left panel) and the position-velocity diagrams for both H$\alpha$ and [\ion{S}{ii}]\,6716\AA\ lines, built from slit 1  (central panel) and slit 2 (right panel). }
\label{fig4}
\end{figure*}

A two-dimensional map of radial velocities  reveals high velocity channel starting from the newly formed knot A and stretching up to knot I (Fig.\,\ref{fig3}).
Lower velocity structures are located around this channel. Of particular interest are knots D and I, where two zones with different radial velocities are observed. In both cases, taking into account the direction of the flow, low-velocity structures are located ahead of higher velocity ones. Integrated values of heliocentric radial velocities in the each knot of the outflow are presented in  Table\ \ref{table}.

It is also noteworthy that our estimates of the radial velocities in the HH~215 jet are rather close to the values, obtained in the work of \citet{caratti} for the near-infrared forbidden emission lines of ionized iron in the spectrum of the source star PV~Cep. As can be seen from their paper, these lines are split to the four components, alike other forbidden emissions in the PV~Cep spectrum \citep[see, e.g.][]{MagMov2001}. We selected from their table A.1 the values of the radial velocities of the  blue-shifted components (HVC-blue) and computed their mean velocity, which turned to be equal to $-$265$\pm$3 km s$^{-1}$. It should be compared with the mean radial velocity of HH~215 knots (excluding the knot G, which is astray from the main flow)   in H$\alpha$ ($-$240$\pm$15 km s$^{-1}$) and [\ion{S}{ii}] ($-$242$\pm$22 km s$^{-1}$) emissions (note, that these last values are corrected to the local standard of rest, which was used in the work of  \citealt{caratti}).

To clarify the positions and values of high and low velocity components we  generated two pseudo-slits based on the data cube, obtained in 2020, along the position angles of 2$^{\circ}$ and 330$^{\circ}$ degrees and width of 2\arcsec. Fig.\,\ref{fig4} shows the monochromatic image in H$\alpha$ emission with marked positions of the pseudo-slits  and position-velocity (PV) diagrams in the H$\alpha$ and [\ion{S}{ii}]\,6716\AA\ lines. We used continuum  subtracted cubes to isolate pure emission and exclude the continuum of the source star and the star, projecting near the knot I. On the PV diagram, corresponding to the first pseudo-slit (PA=2$^{\circ}$) four  knots as well as the bright and wide H$\alpha$ emission from the source are discernible.  The  split of H$\alpha$ emission line in the knot D on two components, with velocities of about $-$255 and $-$180 km s$^{-1}$, should be noted. In contrast, such a split in the [\ion{S}{ii}]\,6716\AA \ emission in the knot D is absent. One should note also that the newly formed knot A is well seen near the bright blueshifted H$\alpha$ emission of the source star, but the pronounced decrease of the radial velocities with distance from the source in this knot is visible only in [\ion{S}{ii}]\,6716\AA.

 The second PV\ diagram, built on the base of the pseudo-slit with PA=330$^{\circ}$, includes five knots. It not only confirms the split of the  H$\alpha$ emission line in the knot D, but also shows similar split in the knot I, with velocities of about $-$245 and $-$185 km s$^{-1}$. These values are very close to those in the knot D. In [\ion{S}{ii}]\,6716\AA\ emission, as in the previous case, lines are not split and show only one  component.

\begin{figure}
\includegraphics[width=10pc]{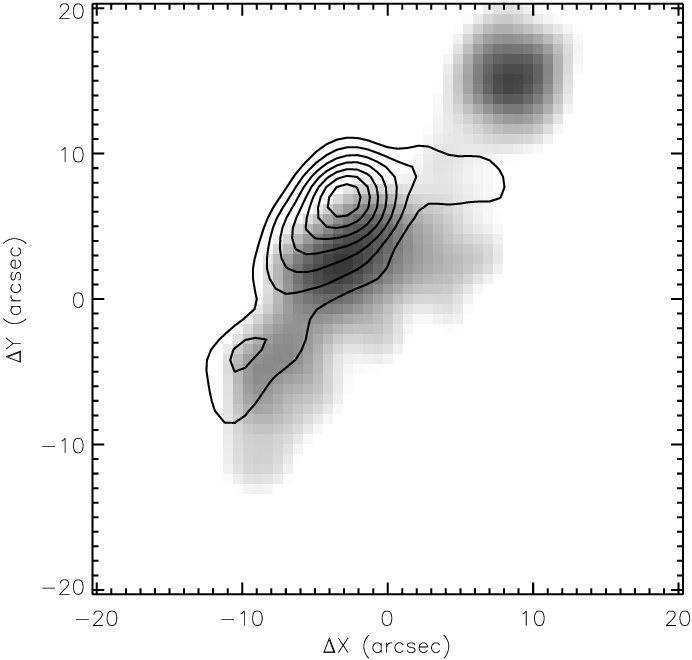} 
\includegraphics[width=10pc]{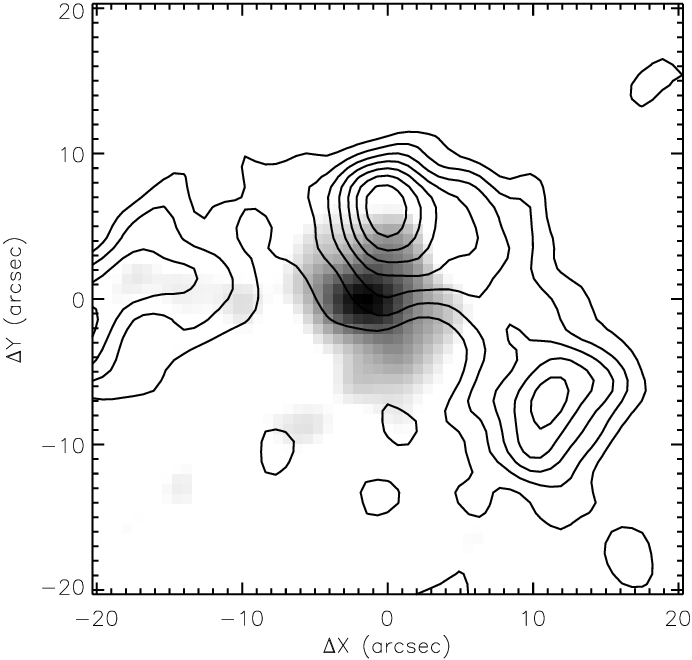} 

\caption{ High (grey scale) and low (isophotes) radial velocity structures in H$\alpha$ emission at knot D (left panel) and knot I (right panel). }
\label{fig5}
\end{figure}

Especially interesting is the fact that the regions with two distinct velocities are spatially separated as well, which has already been observed from the two-dimensional velocity field of the outflow system (Fig.\,\ref{fig3}). To show it even more clearly
we present on the Fig.\ref{fig5} the images of knots D and I, built in channels corresponding to high radial velocity, with the superimposed images in low velocity channels. More precisely, we used high-velocity channels, which correspond to $-$270 km s$^{-1}$ for knot D and $-$255 km s$^{-1}$ for knot I, while low-velocity channels correspond to  $-$175 km s$^{-1}$ for both panels. As one can see, low-velocity structures are located ahead of high-velocity structures for both knots. In the case of the knot I, this effect seems typical for working surfaces, since the high-velocity structure is represented by a compact knot while an arcuate low-velocity structure circles it. 
 
\begin{table*}
\caption{Proper motions and heliocentric radial velocities of knots in the PV~Cep outflow}
\label{table}
\centering
\begin{tabular}{l c c c c c c}
\hline\hline
Knot & Distance\tablefootmark{a}  & V$_{tan}$\tablefootmark{b} & V$_{tan}$ & PA & V$_{r}$ (H$\alpha$) & V$_{r}$ ([\ion{S}{ii}]\,6716\AA)\\  & (arcsec) & (km s$^{-1}$)
&   (arcsec yr$^{-1}$) & (deg) & (km s $^{-1}$) & (km s $^{-1}$) \ \ \ \ \\
\hline
A               &  4.5 &   -   & -  & -   &   $-$252 & $-$237\\
B               &  14.0  &   96 $\pm$ 35 & 0.059  &   328   &   $-$242 & $-$236\\
C              &  19.7 &     -   &   -    &  -& $-$260 & $-$232\\
D  &  25.1 &   205 $\pm$ 40  & 0.124  &   10   &   $-$255 \& $-$180 & $-$281\\
E              &  31.2 &    169 $\pm$ 40   &   0.102    & 343 & $-$280 &    $-$276\\
F              &  39.6  &   240 $\pm$ 35 &  0.145  &   331   &   $-$270 & $-$275\\
G  &  53.7   &   61 $\pm$ 23 & 0.036 &   17 & $-$200 & $-$192\\
H  &  59.2   &  -  & - & -   & $-$230 \\
I  &  61.7   &   125 $\pm$ 30 & 0.075 &  325 & $-$245 \& $-$185 & $-$255\\
\hline
\newline
\end{tabular}
\tablefoot{
\tablefoottext{a}{Distances of knots from the source are measured on 2020 data cube.}
\tablefoottext{b}{Tangential velocities correspond to 350 pc as a distance of the PV~Cep.}
}
\end{table*}

%%%%%%%%%%%%%%%%%%%%%%%%%%%%%%%%%%%%%
\subsection{Proper motions}
%%%%%%%%%%%%%%%%%%%%%%%%%%%%%%%%%%%%

We used restored [\ion{S}{ii}]\,6716\AA\  images, obtained from IFP\ observations in two  epochs to measure proper motions (PM)  of all detectable outflow structures. Unfortunately, we do not have observations in H$\alpha$ during the first epoch, and therefore we were unable to measure proper motion for structures with different radial velocities, which are seen only in  H$\alpha$ emission. As was described above, observations in two different epochs were carried out using different equipment and different detectors, and to compare these data, the images were spatially scaled. We used field stars as reference points. It should be noted that the PM values were measured only for the features, corresponding to the same radial velocity in both epochs, because the inner structures of the PV\ Cep outflow undergo strong morphological changes depending on the radial velocity.
Thus, to achieve better comparison of the morphological details, we created rebinned data cubes for both epochs to bring them to the same velocity steps.
 
\begin{figure}
\includegraphics[width=12pc]{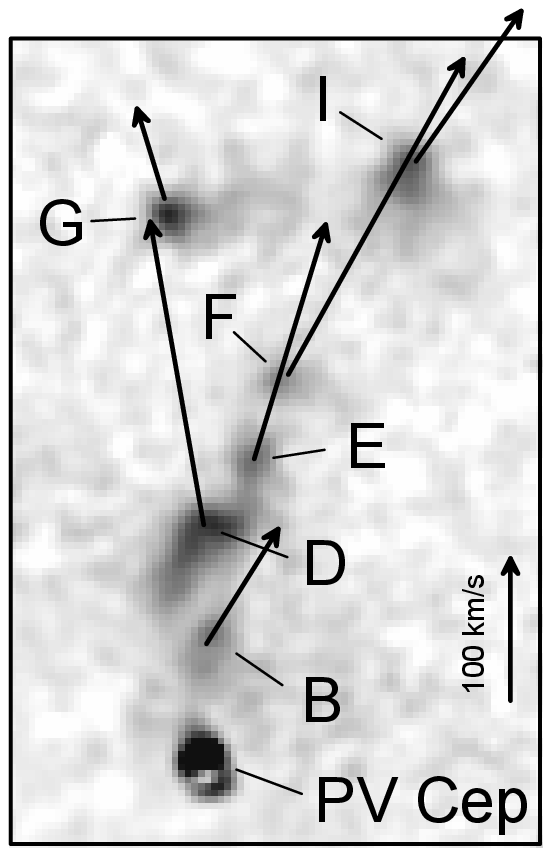}
\caption{ Proper motions of the  knots of HH~215 outflow, shown by vectors. The scale of the vectors is
indicated by the arrow at the right side of the panel.  }
\label{fig6}
\end{figure}

Results are presented in Table\ \ref{table}: the distances for each knot, measured from the central source (PV~Cep), values of tangential velocities (computed for the distance of 350 pc ), PA and also the radial velocities, described in the previous section.
Fig.\,\ref{fig5} shows the PM vectors for the jet knots.
It is obvious that four knots located along the high radial velocity channel have PA\ near 330$^{\circ}$ while others -- near 25$^{\circ}$.

%%%%%%%%%%%%%%%%%%%%%%%%%%%%%%%%%%%%%%%%%%%%%%%%%%%%%%%%%%%%%%%%%%%%%%%%%%%%%%%%
\section{ Discussion and Conclusion }
%%%%%%%%%%%%%%%%%%%%%%%%%%%%%%%%%%%%%%%%%%%%%%%%%%%%%%%%%%%%%%%%%%%%%%%%%%%%%%%%

Before moving on to a detailed consideration of the structure of the flow, it is necessary to clarify its orientation. It should be noted that an obvious discrepancy exists in this question. As can be seen from the studies in the optical range \citep{Reipurth1997,Gomez1997}, rather extended HH~315 flow resembles helical structure but in general is directed to the NNW (northern lobe). Thus, the PA of its axis is somewhat
uncertain, but, in any case, it is close to the symmetry axis of the reflection cometary nebula
near PV~Cep and, besides, it propagates fully in the limits of the northern lobe of the bipolar CO outflow \citep[see][]{Levreault1984,Gomez1997}. Considering the immediate neighborhood of PV~Cep, i.e. HH~215 knots, we see
that they in general follow the same direction, including their proper motion vectors (with exception of knots D and G). 

However, the high spatial resolution millimeter observations \citep{Hamidouche} definitely contradicts to mentioned above  arguments. According to 1.3 mm data, the detected circumstellar disk of PV~Cep has inclination 62\degr\ to the plane of sky and the position angle of the semi-major axis of its ellipse is 297\degr. Therefore, its rotation axis should have PA = 27\degr. This value greatly differs with the mean direction of HH~215 outflow (332\degr\ by PM vectors). The 2.7 mm data, though, do not confirm these estimates, demonstrating nearly circular disk around PV~Cep.   Even more ambiguity arises when we consider the inner parts of the CO molecular outflow (mapped in the same work), which seems to propagate at first orthogonally to the circumstellar disk, but after moving away to a distance of about 25\arcsec\ it turns clockwise probably becoming more aligned with HH~313 flow and its CO counterpart. However, it seems that on these maps we probably see the most dense edges of the extended CO outflow  (see maps of \citealt{AG2002}), and due to the very small beam the
main part of the outflow emission was filtered away.
Thus, we tend to assume as 
more credible values 332\degr\ for PA and 30\degr\ for $i$ (see next paragraph). 

Presently, having more or less complete picture of the 3D kinematics of the HH~215 knots, we can directly derive the spatial orientation of the ionized flow. It is quite logical to use for the computation the knots, located near the high-velocity channel,   PM vectors of which  also have nearly the same direction, -- namely B, E, F and I. Deriving the mean values of V$_{r}$ and V$_{t}$ for them from the values from Table \ref{table}, we get  
i $\approx$ 30\degr$\pm$ 5\degr\ for inclination angle between the flow axis and the line of sight. Thus, the inclination of the circumstellar disk of PV~Cep to the plane of the sky should be of same value. 

The close match of the radial velocities of the HH~215 knots and of the jet in the immediate environment of PV~Cep, described in the previous section, should be considered as the confirmation that the flow as whole starts near the circumstellar disk  and does not undergo noticeable changes in direction. Nevertheless, one should reconsider the full velocity of the northern branch of the  ionized flow, which was estimated by \citet{caratti} as about 600 km s$^{-1}$ (however, one should keep in mind that this estimation was based not only on their measured radial velocities, but also on the inclination angle derived from \citealt{Hamidouche}).
Taking into account our new data about tangential velocities we get about 300 km s$^{-1}$, what is quite typical for jet velocities and shows, that our value for the inclination angle of the flow is much more consistent and the previous speed of the flow was grossly overestimated. Besides, if the inclination, estimated by \citet{Hamidouche}, were correct, the PM, corresponding to tangential velocity of the northern branch of the outflow, would be much greater (about 500 km s$^{-1}$), definitely exceeding the observation errors of our measurements.  

The morphology of the HH~215 outflow seems somewhat unusual in the sense that it is divided into two parts: the wiggling jet and the preceding arcuate structure with an opening angle of about 40$^{\circ}$ toward the source.
 It is possible to deproject the observed emission structures and to get the morphology of the flow how it would be visible
if the flow axis coincided with the plane of the sky. This can be obtained by extending the dimensions along the axis by a factor
of 1/sin 30$^{\circ}$ = 2. The result is shown in Fig.\ref{fig7}. As it can be seen, the opening angle of the bow-shaped structure after the deprojecting becomes about 15$^{\circ}$, which is more typical for terminal working surfaces. The total length of HH~215 outflow will be about 0.2 pc, and the total length of {the bipolar outflow from PV~Cep (HH~315 + HH~215), estimated  by \citet{Reipurth1997} for the distance of 500 pc as 2.6 pc in projection, will become for the newly estimated distance of PV~Cep 350 pc only about 3.6 pc, assuming that it more or less keeps the same inclination angle.

As the one of the most important results of this study we consider the discovery of a newly emerged emission knot (A) near
the source in 2020, which was not visible in 2003.  The formation of new knots in stellar jets is a well-known phenomenon; see, e.g.,  \object{HH~1} jet \citep{Hartigan2011} and \object{HH~34} \citep{Reipurth2002}. This can be considered as an indication that the sources continue to eject the matter episodically. Several new knots in the jet of Z~CMa, probably connected with episodic outbursts, were detected by \citet{Whelan}.

The projected distance
of this newly formed knot from the source is about 4\arcsec, which  on the distance of PV~Cep (350 pc) is equal to $\approx 1400$ a.u. Though the PM of knot A cannot be measured directly since it was not visible in the first epoch, we can assume for the knot A the mean tangential velocity of B, F, E and I knots:
$\approx$ 145 km s$^{-1}$. Then the kinematic age of the knot A can be derived as about 46 years. Considering the lightcurve of PV~Cep,  this age near perfectly corresponds to the period of the maximal brightness of the source (1976-1978), when this star was discovered \citep{AMMM2021}. This  makes the case of PV~Cep the one of first observational proofs of the probable connection between the formation of HH jet knotty structures and the outbursts of the central source \citep{Reipurth1997}. One should note, of course, that in the work of \citet{caratti} yet another knot was found just near the PV~Cep star (1\arcsec) in the southern branch
of the outflow. However, the new estimates of the star distance and of the jet inclination allow to infer from its kinematic age the 1995 as the probable date of its formation;
unfortunately we do not know about any photometric observations in that period.

Another important step in the reconstruction of the morphology of HH~215
was the revealing its complexity, consisting of wiggling HH flow and a frontal huge bow-shaped
structure.  Since the first epoch observations were performed when PV~Cep star, as well as its associated reflection nebula were at its maximal
brightness, it was possible to combine the reconstructed image of the nebula in the
continuum and the data  in H$\alpha$  to better understand the positions of emission knots and the orientation  of the flow against the
cone-shape reflection nebula.
During the second epoch observations for the first
time this flow was studied in H$\alpha$ and [\ion{S}{ii}]\,6716\AA\, emissions. 

In this work the full kinematics of the HH~215 system
is described, also for the first time, and the correct value of the jet inclination was obtained. Besides, new data about the distance, jet velocity and the disc inclination should lead to revision of the several stellar parameters, for example of luminosity and mass loss rate of PV Cep, though such reanalysis is beyond the scope of the present paper. In any case, the star will become more similar to EXors.  

The revealing of the high-velocity channel passing through the center axis of the outflow and coinciding with the trajectory of the giant outflow \citep{Reipurth1997}  should be mentioned.
As expected, it is directed along the axis of the reflection nebula, once more proving the generic connection between the outflow and circumstellar disk, which is responsible for the formation of fan-shaped nebulae. Similar cases of entrained  jets are known for many years \cite[see, e.g.][]{HMHC}. 

Detailed studies revealed two knots (D and I)
whose H$\alpha$ emission exhibits two regions which are both spatially and kinematically
separated. The case of knot D appears more close to the standard picture, where a planar shock, embedded within a bow shock, is formed by oblique shock when wiggling jets encounter cavity walls, because the high and low velocity components exhibit similar morphology. According to  \cite{Hartigan2011}, such type of interaction must be common in stellar jets, where ejection angle variations of several degrees within a given jet  are usual.

The situation with knot I is different: the high velocity component represents a compact structure, while the low velocity one is more extended and has a bow-shaped morphology. In our opinion the high- and low-velocity structures in the knot I represent bow-shock and reverse shock in the
internal working surfaces of the episodic outflow from PV~Cep. Such configuration of shocks is similar to the case of \object{HL~Tau} jet \citep{Mov2012}.  

Two epoch observations give us possibility to measure PM of individual knots in HH~215 system. In general, knots have a large PM value, which varies greatly from one knot to another.
Low PM is indicated in the huge bow-shape structure, where outflow material probably collides with the matter, previously ejected with lower velocity. Position angles of PM vectors vary from knot to knot. Knots located along the axis of the high-velocity channel have a position angle coinciding with its axis (about 325$^{\circ}$);  other ones have completely different value (about 25$^{\circ}$), which supports the idea that those knots are formed by oblique shocks.
However, we have not found a single similar example in the literature, where PM of the knots belonging to the same jet, show so large differences between each other, with exception of the cases of significant interaction of the flows between each other or with surrounding cavity walls \citep{Mov2007,Lopez, Massi}. The evidences of such interaction in the HH~215 flow are not found yet.

\begin{figure}
\includegraphics[width=8pc]{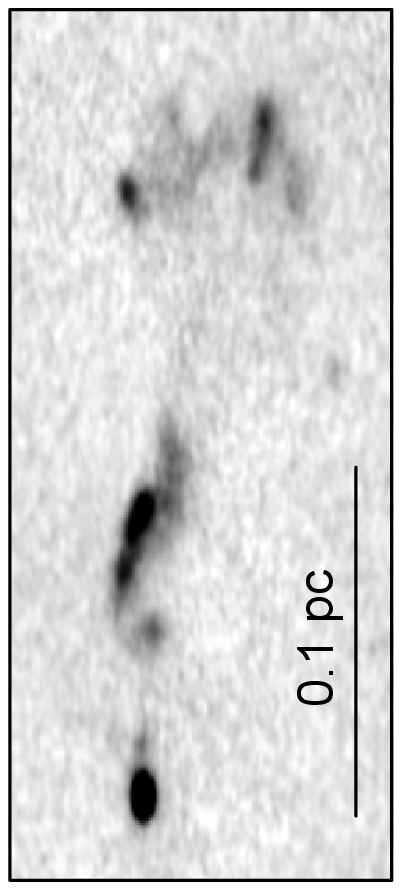}
\centering
\caption{Gray scale image of the HH 215 outflow in H$\alpha$,  deprojected   from the Fig. 2a  with  a factor of 1/sini=2.    }
\label{fig7}
\end{figure}

\begin{acknowledgements}

We wish to thank the referee, Dr. Fabrizio Massi, for his very encouraging insights and comments. We are grateful to Dr. Dmitry Oparin and Dr. Roman Uklein who preformed SCORPIO-2 observations in 2020-21. We obtained  the observed data on the unique scientific facility "Big Telescope Alt-azimuthal"  of SAO RAS as well as made data reduction with the financial support of grant No075-15-2022-262 (13.MNPMU.21.0003) of the Ministry of Science and Higher Education of the Russian Federation.  This work was supported by the RA MES 
State Committee of Science, in the frames of the research project number 21T-1C031.

\end{acknowledgements}
   
%%%%%%%%%%%%%%%%%%%%%%%%%%%%%%%%%%%%%%%%%%%%%%%%%%%%%%%%%%%%%%%%%%%%%%%%%%%%%%%%

\end{document}